\documentclass[preprint,12pt, a4paper]{elsarticle}
\usepackage{graphicx}
\usepackage{subcaption}
\usepackage{amssymb}
\usepackage{hyperref}
\usepackage{cleveref}
\journal{SoftwareX}

\begin{document}
\renewcommand{\labelenumii}{\arabic{enumi}.\arabic{enumii}}

\begin{frontmatter}

\title{\texttt{Magboltz-GUI}: a Python-based graphical user interface for \texttt{Magboltz}}

\tnotetext[ded]{Dedicated to the memory of Stephen Francis Biagi (1949--2025)}

\author[label1]{Michele Renda}
\author[label1]{Dan Andrei Ciubotaru}
\author[label1,label2]{Călin Alexa}
\address[label1]{Particle Physics Department, IFIN-HH, 077125 Magurele, RO}
\address[label2]{calin.alexa@cern.ch}

\begin{abstract}
\texttt{Magboltz}\cite{magboltz} is widely used to compute electron transport properties in gas mixtures for detector applications. Its text-based workflow, however, can be a barrier for routine use, especially for users who are not already familiar with the program. We present \texttt{Magboltz-GUI}, a Python-based graphical user interface for defining gas mixtures, configuring simulation parameters, running \texttt{Magboltz}, and visualizing or exporting the resulting. The tool is designed as a lightweight frontend for common tasks in research and teaching environments involving gaseous detectors, including micropattern technologies such as Micromegas \cite{micromegas}. This paper describes the software implementation, main interface components, and its availability as an open-source distributed package via Python tools.
\end{abstract}


\end{frontmatter}

\section*{Metadata}

\begin{table}[!h]
\footnotesize
\begin{tabular}{|l|p{6.5cm}|p{5.5cm}|}
\hline
\textbf{Nr.} & \textbf{Code metadata description} & \textbf{Metadata} \\
\hline
C1 & Current code version & v1 \\
\hline
C2 & Permanent link to code/repository used for this code version & \url{https://gitlab.com/micrenda/magboltz-gui} \\
\hline
C3 & Permanent link to Reproducible Capsule & N/A \\
\hline
C4 & Legal Code License & MIT License \\
\hline
C5 & Code versioning system used & git \\
\hline
C6 & Software code languages, tools, and services used & Python, Qt \\
\hline
C7 & Compilation requirements, operating environments \& dependencies & Python $\ge$ 3.11, Qt \\
\hline
C8 & If available Link to developer documentation/manual & N/A \\
\hline
C9 & Support email for questions & michele.renda@cern.ch \\
\hline
\end{tabular}
\caption{Code metadata}
\label{codeMetadata}
\end{table}

\section{Motivation and significance}

\texttt{Magboltz}\cite{magboltz} is a widely used tool for low-temperature plasma and gaseous-detector simulations. Developed by Stephen Biagi at CERN, it computes electron drift and related transport properties in gas mixtures under static electric and magnetic fields by modeling elastic and inelastic collision processes.

Because of these capabilities, \texttt{Magboltz} is routinely used in the study and design of gaseous detectors, including time projection chambers (TPCs), drift chambers, Micromegas \cite{micromegas}, and gas electron multipliers (GEMs) \cite{gem}. It is also commonly used together with larger simulation environments such as \texttt{Garfield++} \cite{garfieldpp} and \texttt{Geant4} \cite{geant4}.

Although \texttt{Magboltz} is scientifically well established, its traditional workflow is based on manually prepared text input cards. This approach is flexible, but it can be inconvenient for routine use and unfriendly to new or occasional users. Small input mistakes may also lead to failed runs or make debugging more difficult.

The goal of \texttt{Magboltz-GUI} is simple: to make routine interaction with \texttt{Magboltz} more convenient without changing the underlying simulation code. The GUI provides a graphical way to define gas mixtures, edit input parameters, run simulations, and inspect or export the resulting outputs. It can also launch multiple runs in parallel, which is useful for repeated runs and simple parameter scans.

The tool is intended both for practical day-to-day use and for teaching or introductory work, where a graphical frontend can make the program easier to approach. The remainder of this paper is organized as follows: Section~2 describes the implementation; Section~3 presents installation instructions; Section~4 describes the interface; and Section~5 summarizes the conclusions and possible future developments.

\section{Implementation}
\label{sec:implementation}

The tool is implemented in Python~3.11 using the Qt~6 bindings for Python \cite{qtforpython}. This choice allows the application to run on the main desktop operating systems and keeps packaging straightforward. The software is distributed through the Python Package Index (PyPI) \cite{magboltzGuiPypi}, so installation and updates can be handled with standard Python tools.

The GUI does not modify the \texttt{Magboltz} source code \cite{magboltzCern}. It interacts with \texttt{Magboltz} through standard input and output (\textit{stdin}/\textit{stdout}), so it can be used with an existing local installation. By default, the GUI uses the executable available on the system \texttt{PATH}, but a custom path can also be set manually.

A local \texttt{Magboltz} installation is required only when simulations are executed from within the GUI. The application can also be used simply to prepare input cards, which can later be run separately in batch mode or in other workflows.

The only mandatory runtime dependencies are Python~3.11 or newer and the Qt~6 Python bindings. When \texttt{Magboltz} is installed locally, the recommended setup is to make the executable available on the system \texttt{PATH}; otherwise, its location can be configured in the application settings.

\section{Installation}
\label{sec:installation}

Two installation methods are provided.

\paragraph{Installation with bundled Qt dependencies}

The simplest option is to install the package with the Qt extra from PyPI:
\begin{verbatim}
pipx install "magboltz-gui[qt]"
\end{verbatim}

This approach works well on macOS and on most Linux distributions. Its main drawbacks are the larger download size, due to the bundled Qt dependencies, and occasional rendering artifacts in some Linux environments.

\paragraph{Installation using a system Qt environment}

Alternatively, users can rely on an existing Qt installation available in the system site-packages:
\begin{verbatim}
pipx install magboltz-gui --system-site-packages
\end{verbatim}

\paragraph{Launching the application}

After installation, the GUI can be started with:
\begin{verbatim}
magboltz-gui
\end{verbatim}

The package requires Python $\ge$ 3.11 and a graphical desktop environment. We recommend using \texttt{pipx} so that dependencies remain isolated while the \texttt{magboltz-gui} command is still available system-wide. A local \texttt{Magboltz} installation is needed only if simulations are run from within the GUI; otherwise, the application can still be used to prepare input cards.

\section{Interface}

The interface is intended to remain simple while preserving a workflow familiar to existing \texttt{Magboltz} users.

The first tab contains the full set of input parameters. Units are shown directly in the interface, and short descriptions are available through tooltips. The lower part of the same tab is used to define gas mixtures, with support for up to six gases, as allowed by \texttt{Magboltz}. A pie chart of the mixture composition is generated with \texttt{matplotlib} and can be exported.

\begin{figure}[h]
	\centering
	\includegraphics[width=\columnwidth]{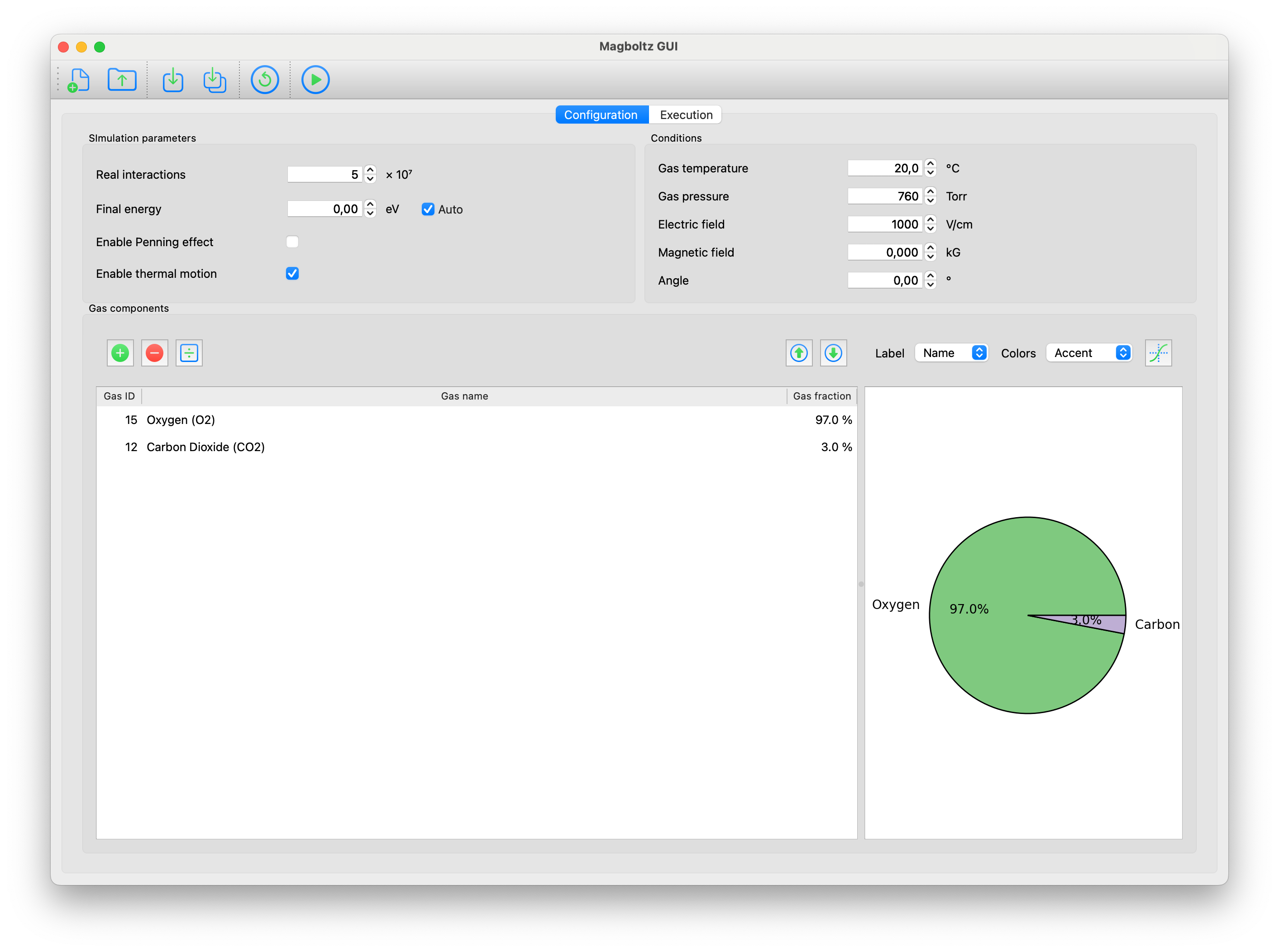}
	\caption{Main window of the \texttt{Magboltz-GUI} application.}
	\label{mai-window}
\end{figure}

The second tab is used to run \texttt{Magboltz} directly from the GUI and to export the resulting outputs in CSV format.

\begin{figure}[h]
	\centering
    \includegraphics[width=0.47\textwidth]{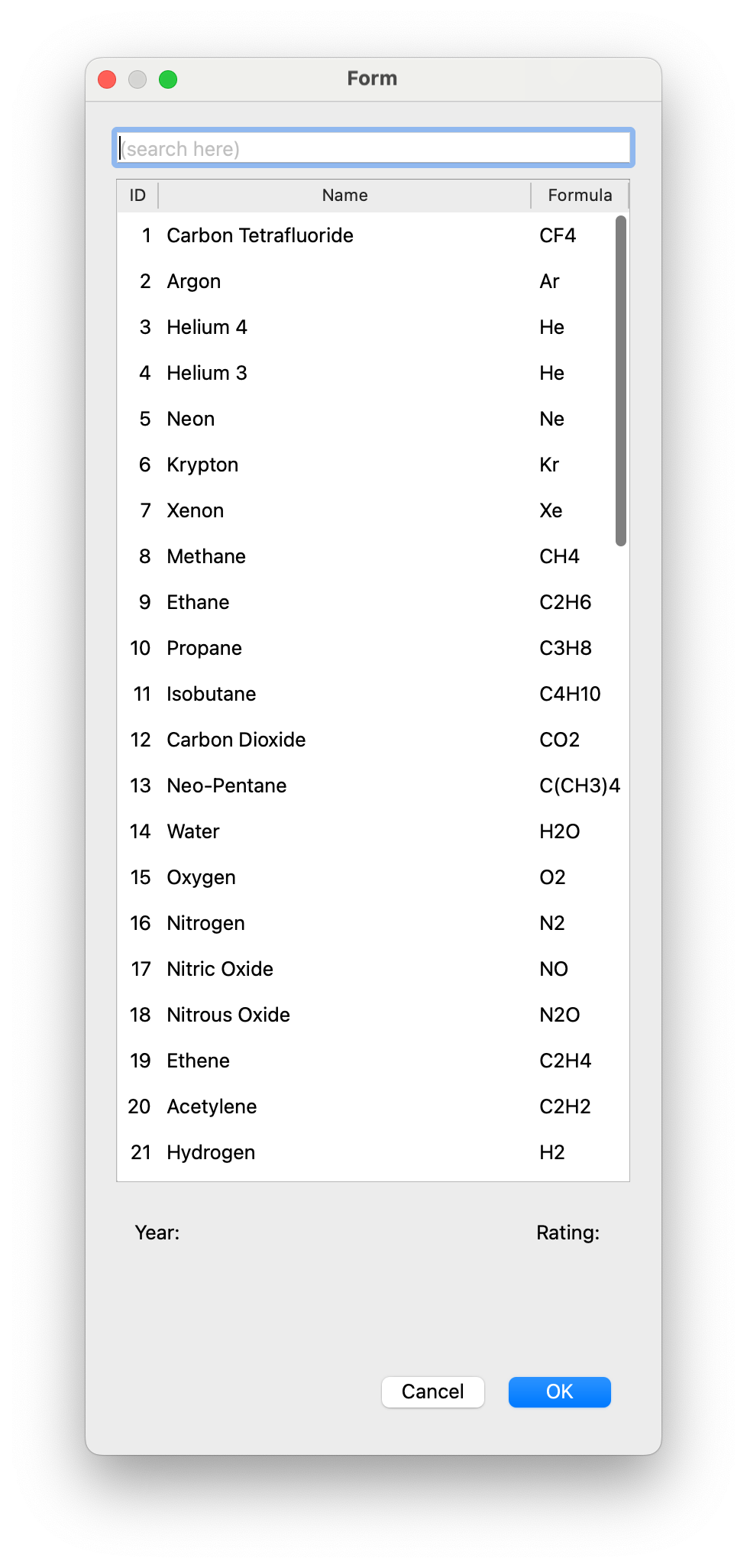}
	\caption{Gas selection window.}
	\label{main-window}
\end{figure}

\subsection{Export simulation data}

\begin{figure}[h]
	\centering
    \includegraphics[width=.75\textwidth]{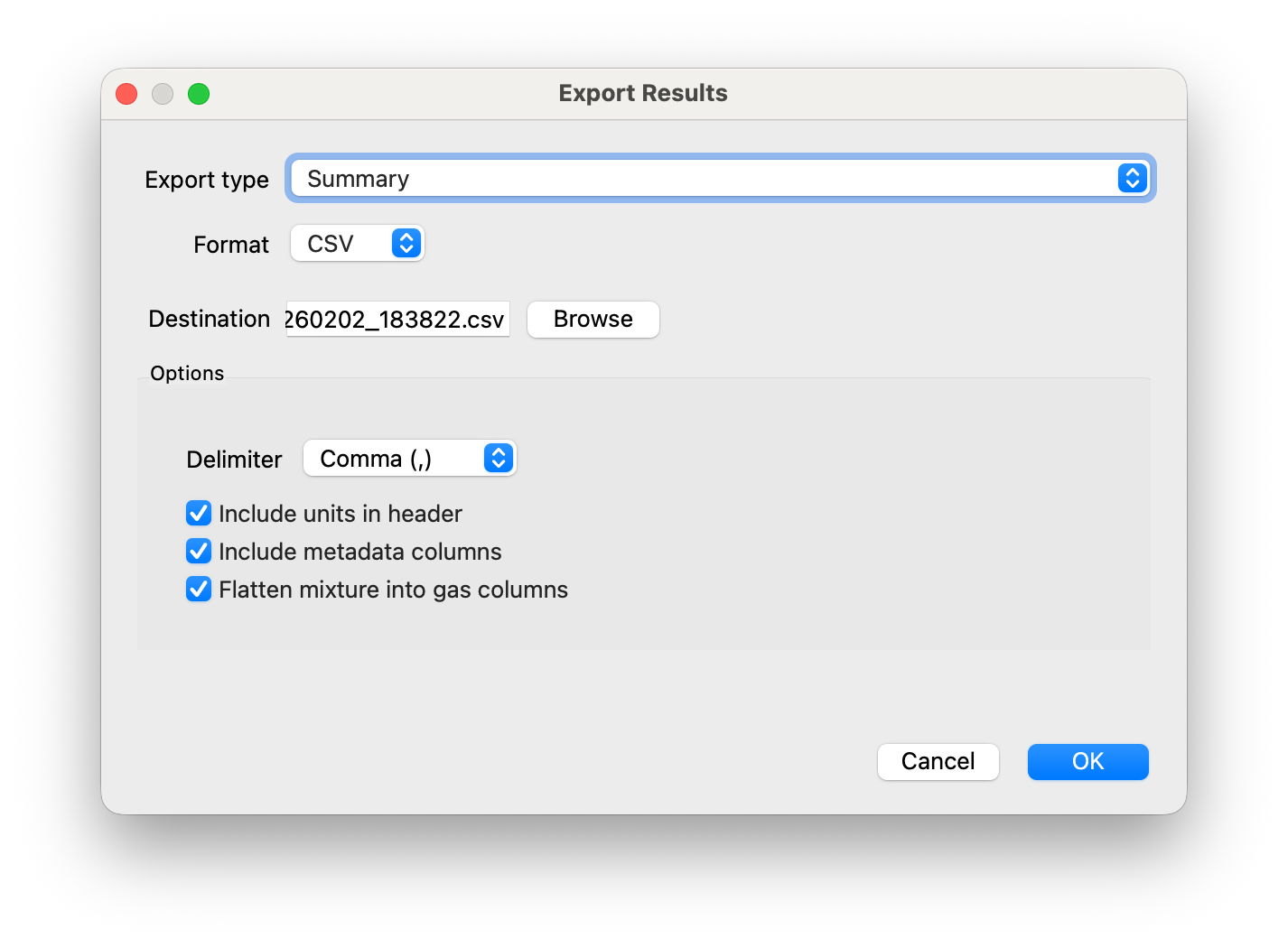}
	\caption{\textit{Export Results} dialog showing the available export types, output formats, and data customization options.}
	\label{fig:export-window}
\end{figure}

The \textit{Export Results} dialog (\cref{fig:export-window}) provides several predefined export types, including summary outputs, convergence tables, energy distributions, collision frequencies by gas and process, and full-run archives. Results can be written in CSV, JSON, or XML format, with optional inclusion of units, metadata, and expanded gas-mixture information. This makes it easier to reuse \texttt{Magboltz} outputs in external analysis tools.

\subsection{Visualizing simulation results}

\begin{figure}[h]
    \centering
    \begin{subfigure}{0.45\textwidth}
        \centering
        \includegraphics[width=\linewidth]{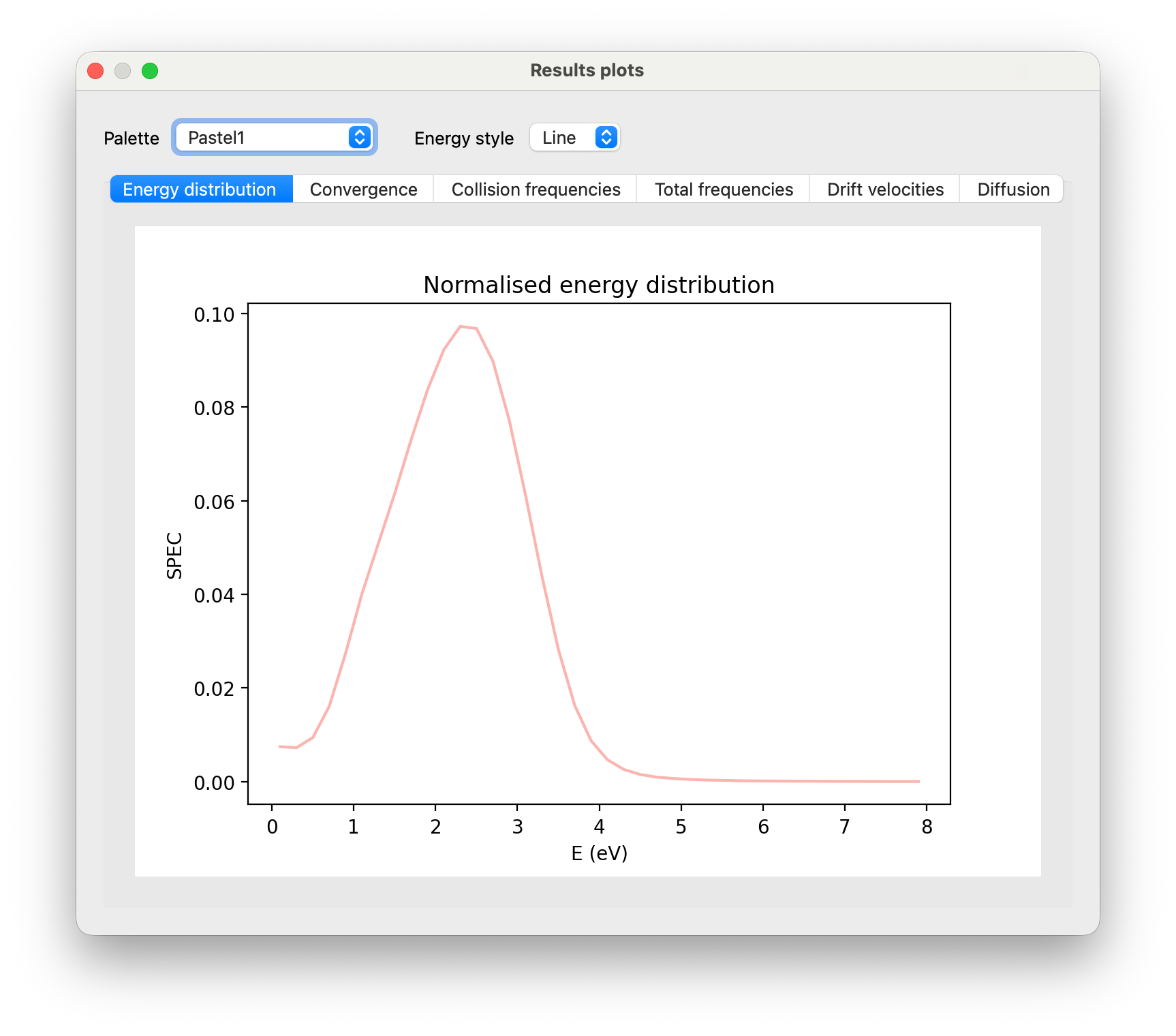}
        \caption{Energy distribution graph.}
    \end{subfigure}
    \hfill
    \begin{subfigure}{0.45\textwidth}
        \centering
        \includegraphics[width=\linewidth]{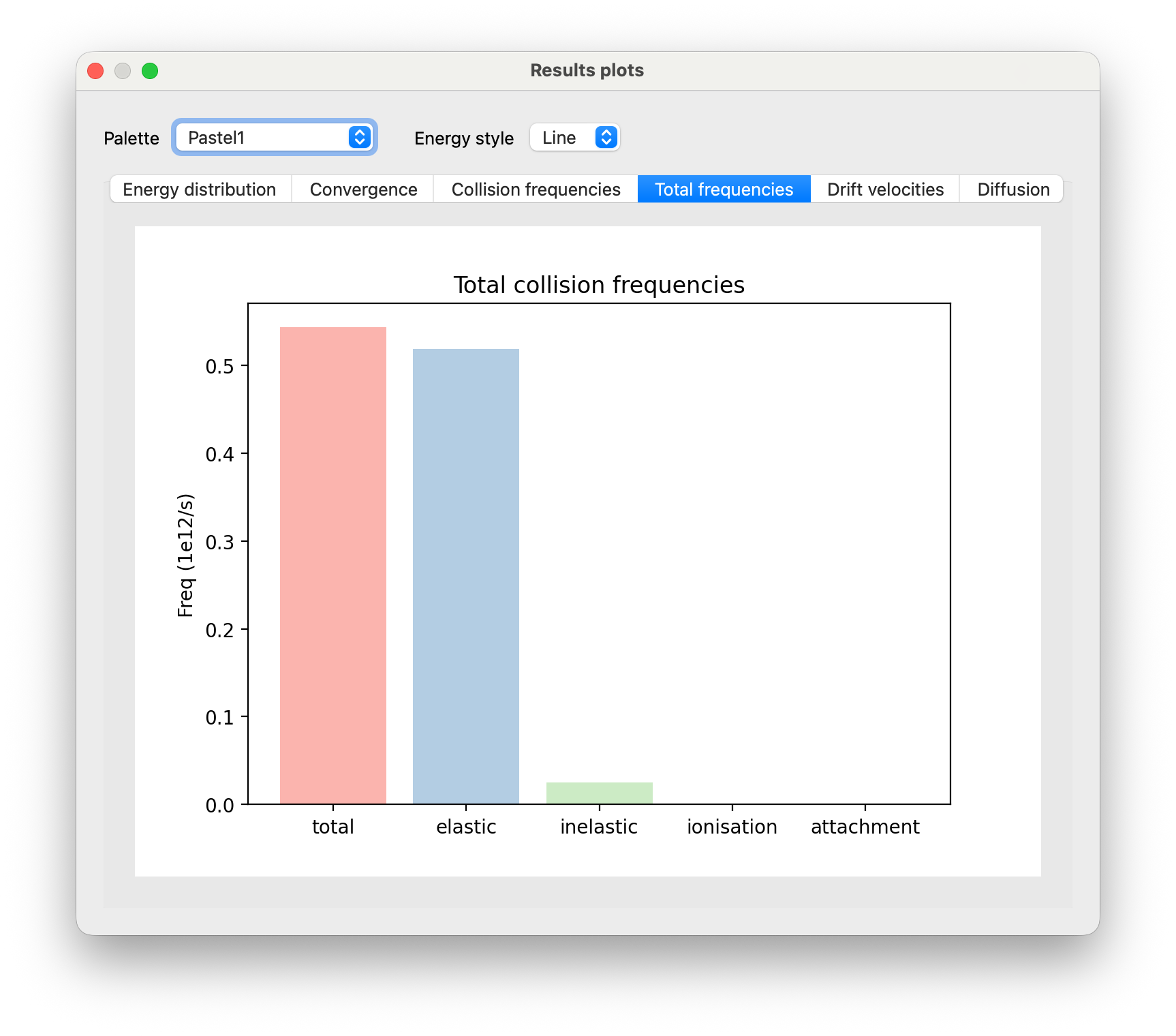}
        \caption{Total collision frequencies graph.}
    \end{subfigure}

    \medskip

    \begin{subfigure}{0.45\textwidth}
        \centering
        \includegraphics[width=\linewidth]{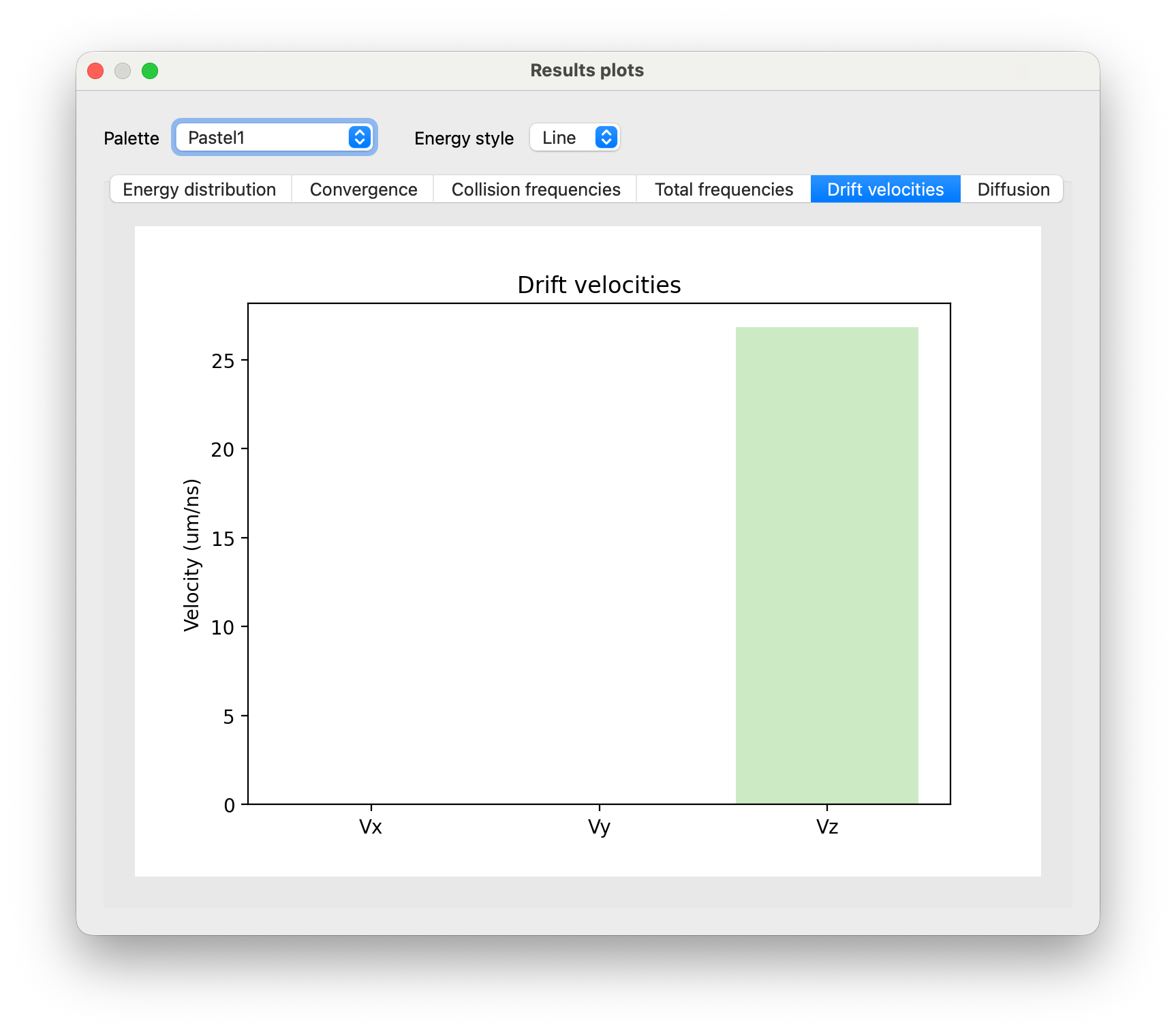}
        \caption{Drift velocities graph.}
    \end{subfigure}
    \hfill
    \begin{subfigure}{0.45\textwidth}
        \centering
        \includegraphics[width=\linewidth]{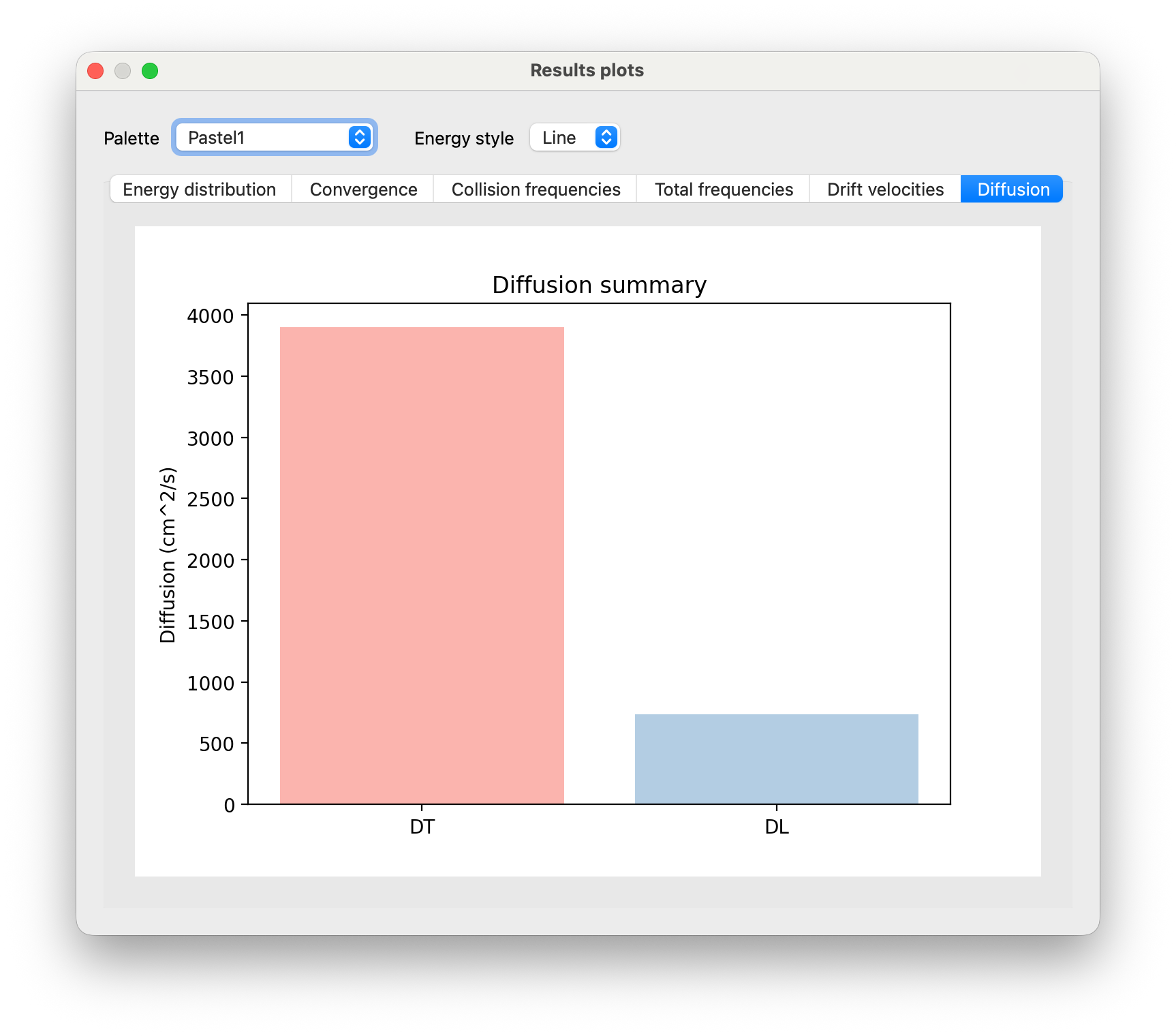}
        \caption{Diffusion distribution graph.}
    \end{subfigure}

    \caption{\textit{Results plots} window showing several views of \texttt{Magboltz} simulation outputs, with selectable result types and customizable plot styles.}
    \label{fig:graphs}
\end{figure}

Figure \labelcref{fig:graphs} shows the \textit{Results plots} window. It provides graphical views of several \texttt{Magboltz} outputs, including energy distributions, convergence behavior, collision frequencies, drift velocities, and diffusion coefficients. Plot appearance can be adjusted through selectable color palettes and plot styles.

\subsection{Availability \& Licensing}

The software is released under the MIT License, permitting use, modification, and redistribution. The source code is publicly available in a GitLab repository, and installable releases are distributed through PyPI \cite{magboltzGuiGitlab,magboltzGuiPypi}.

\section{Conclusions}
\label{sec:conclusion}

\texttt{Magboltz-GUI} provides a lightweight graphical frontend for routine interaction with \texttt{Magboltz}. It facilitates common tasks such as preparing input cards, defining gas mixtures, launching runs, and viewing or exporting outputs more convenient, while remaining compatible with existing \texttt{Magboltz} installations.

The tool was presented at the DRD1 \cite{DRD1} WG4 meeting on 25 February 2026, where feedback from the community was collected and has informed subsequent improvements.

The current implementation focuses on usability and workflow simplification rather than modifications of the underlying simulation model. Future developments will include enhancements to the interface, extended plotting and export capabilities, and further improvements guided by user feedback.

Its modular Python/Qt6 design facilitates reuse and extension in related workflows and detector-simulation studies.

Overall, the \texttt{Magboltz-GUI} provides a practical layer on top of \texttt{Magboltz}, lowering the barrier to entry for new users and enabling more efficient and reproducible workflows for routine studies.

The software is released under the MIT License and is available via PyPI and the public repository.

\section*{Acknowledgements}
We thank the \texttt{Magboltz} user community for discussions and feedback during the development of \texttt{Magboltz-GUI}. We also acknowledge the broader gaseous-detector and low-temperature plasma communities, whose open software ecosystem made this work possible.

\subsection*{Funding}
This work was supported by IFIN-HH under Contract No. \texttt{PN-23210104} with the Romanian Ministry of Education and Research. The continued development of this software is guided by the work of the authors and by feedback from users in the research community.

\subsection*{In Memory of Stephen F. Biagi}
We dedicate this work to the memory of \textbf{Stephen F. Biagi} (1949--2025), whose contributions to the modeling of electron transport in gases have had a lasting impact on detector physics. Through the development and long-term maintenance of \texttt{Magboltz}, he provided the community with a tool that has supported generations of researchers working on gaseous detectors. This interface is offered as a small practical tribute to that legacy.

\bibliographystyle{elsarticle-num}
\bibliography{cas-refs}

\end{document}